\documentclass[a4paper,UKenglish,cleveref,autoref,thm-restate,authorcolumns,cleveref]{lipics-v2021}
\usepackage[abbreviations]{foreign}
\usepackage{xspace}
\usepackage[utf8]{inputenc}
\usepackage[T1]{fontenc}
\usepackage[prologue,usenames,table,dvipsnames,x11names]{xcolor}
\usepackage{amsmath,amsfonts}
\usepackage{algorithmic}
\usepackage{graphicx}
\usepackage{textcomp}
\usepackage{listings}
\usepackage{url}
\usepackage{pifont}
\usepackage{booktabs} \usepackage{multirow}
\usepackage{tabularx}
\usepackage{makecell}
\usepackage{siunitx}
\usepackage{float}
\usepackage[export]{adjustbox}
\usepackage{eso-pic}
\usepackage[most]{tcolorbox}
\usepackage{calc}

\usepackage[shadow=false]{pgf-umlsd}
\newcommand{\newthreadshift}[3][0.2]{
  \newinst[#1]{#2}{#3}
  \stepcounter{threadnum}
  \node[below of=inst\theinstnum,node distance=0.8cm] (thread\thethreadnum) {};
  \tikzstyle{threadcolor\thethreadnum}=[fill=gray!30]
  \tikzstyle{instcolor#2}=[fill=gray!30]
}

\nolinenumbers

\def\wasm{Wasm\xspace}
\def\Wasm{\wasm}
\def\wasmSGX{\wasm-SGX\xspace}

\def\pubsub{pub/sub\xspace}
\def\Pubsub{Pub/sub\xspace}
\def\palaemon{\textls[-50]{\rm\textsc{Pal{\ae}mon}}\xspace}
\def\attreq{\textbf{AttReq}\xspace}
\def\attclient{\textbf{AttClient}\xspace}
\def\attserver{\textbf{AttServer}\xspace}
\def\clienthello{\textit{ClientHello}\xspace}
\def\serverhello{\textit{ServerHello}\xspace}
\newcommand{\Ding}[1]{\raisebox{-0.8pt}{\ding{\the\numexpr #1 + 191}}}
\newcommand{\DingBlack}[1]{\raisebox{-0.8pt}{\ding{\the\numexpr #1 + 201}}}

\title{A Holistic Approach for Trustworthy Distributed Systems with WebAssembly and TEEs}
\titlerunning{\resizebox{398pt}{!}{A Holistic Approach for Trustworthy Distributed Systems with WebAssembly and TEEs}} 

\author{Jämes Ménétrey}{University of Neuchâtel, Switzerland}{james.menetrey@unine.ch}{https://orcid.org/0000-0003-2470-2827}{}
\author{Aeneas Grüter}{University of Bern, Switzerland}{}{https://orcid.org/0009-0009-2299-3393}{}
\author{Peterson Yuhala}{University of Neuchâtel, Switzerland}{peterson.yuhala@unine.ch}{https://orcid.org/0000-0002-3371-9228}{}
\author{Julius Oeftiger}{University of Bern, Switzerland}{}{https://orcid.org/0009-0002-3416-1198}{}
\author{Pascal Felber}{University of Neuchâtel, Switzerland}{pascal.felber@unine.ch}{https://orcid.org/0000-0003-1574-6721}{}
\author{Marcelo Pasin}{University of Neuchâtel, Switzerland}{marcelo.pasin@unine.ch}{https://orcid.org/0000-0002-3064-5315}{}
\author{Valerio Schiavoni}{University of Neuchâtel, Switzerland}{valerio.schiavoni@unine.ch}{https://orcid.org/0000-0003-1493-6603}{}

\authorrunning{J. Ménétrey, A. Grüter, P. Yuhala, J. Oeftiger, P. Felber, M. Pasin, and V. Schiavoni} 

\Copyright{Jämes Ménétrey, Aeneas Grüter, Peterson Yuhala, Julius Oeftiger, Pascal Felber, Marcelo Pasin, and Valerio Schiavoni}

\ccsdesc[300]{Security and privacy}
\ccsdesc[500]{Security and privacy~Distributed systems security}
\ccsdesc[500]{Security and privacy~Trusted computing}
\ccsdesc[300]{Computer systems organization~Dependable and fault-tolerant systems and networks}

\keywords{Publish/Subscribe, WebAssembly, Attestation, TLS, Trusted Execution Environment, Cloud-Edge Continuum} 

\category{} 

\relatedversion{} 

\supplementdetails{Software}{https://github.com/JamesMenetrey/unine-opodis2023} 

\funding{This publication incorporates results from the VEDLIoT project, which received funding from the European Union's Horizon 2020 research and innovation programme under grant agreement No 957197.}

\EventEditors{Alysson Bessani, Xavier D\'efago, Yukiko Yamauchi, and Junya Nakamura}
\EventNoEds{4}
\EventLongTitle{27th International Conference on Principles of Distributed Systems (OPODIS 2023)}
\EventShortTitle{OPODIS 2023}
\EventAcronym{OPODIS}
\EventYear{2023}
\EventDate{December 6--8, 2023}
\EventLocation{Tokyo, Japan}
\EventLogo{}
\SeriesVolume{286}
\ArticleNo{23}

\newcolumntype{R}{>{\raggedleft\arraybackslash}X}
\newcolumntype{P}[1]{>{\raggedleft\arraybackslash}p{#1}}

\definecolor{prioritycolor}{HTML}{969bce}
\definecolor{darkgray}{HTML}{262626}
\newcommand*{\priority}[2]{\raisebox{-1pt}{\rule{0pt}{11pt}\begin{tikzpicture}[scale=0.15]\color{#2}
  \draw[line width=0.3mm] (0,0) circle (1);
  \fill[fill opacity=1,fill=#2] (0,0) -- (90:1) arc (90:90-#1*3.6:1) -- cycle;
  \end{tikzpicture}}}
\newcommand*{\circlequestionmark}{\raisebox{-0pt}{\rule{0pt}{11pt}\color{darkgray}
    \fontsize{10pt}{12pt}\selectfont
    \faQuestionCircle}}
\newcommand{\compfull}{\priority{100}{ForestGreen}}
\newcommand{\comppart}{\priority{50}{YellowOrange}}
\newcommand{\compnone}{\priority{0}{BrickRed}}
\newcommand{\compunknown}{\circlequestionmark}

\newcommand{\benchmarkRoundtripWasmSGX}{\qty{22}{\micro\second}\xspace}
\newcommand{\benchmarkNewConnectionsWasmVsNative}{2.57$\times$\xspace}
\newcommand{\benchmarkNewConnectionsWasmSGXVsNative}{4.72$\times$\xspace}
\newcommand{\benchmarkNewConnectionsWithAttestationECDH}{1.87$\times$\xspace}
\newcommand{\benchmarkNewConnectionsWithAttestationPSK}{4.33$\times$\xspace}
\newcommand{\benchmarkThroughputWasmVsNative}{1.30$\times$\xspace}
\newcommand{\benchmarkThroughputWasmSGXVsNative}{1.55$\times$\xspace}
\newcommand{\benchmarkScalingPublishersWasmVsNative}{1.31$\times$\xspace}
\newcommand{\benchmarkScalingPublishersWasmSGXVsNative}{1.56$\times$\xspace}

\makeatletter
\begin{document}

\maketitle

\begin{abstract}
    Publish/subscribe systems play a key role in enabling communication between numerous devices in distributed and large-scale architectures.
    While widely adopted, securing such systems often trades portability for additional integrity and attestation guarantees.
    Trusted Execution Environments (TEEs) offer a potential solution with enclaves to enhance security and trust.
    However, application development for TEEs is complex, and many existing solutions are tied to specific TEE architectures, limiting adaptability.
    Current communication protocols also inadequately manage attestation proofs or expose essential attestation information.
    This paper introduces a novel approach using WebAssembly to address these issues, a key enabling technology nowadays capturing academia and industry attention.
    We present the design of a portable and fully attested publish/subscribe middleware system as a holistic approach for trustworthy and distributed communication between various systems.
    Based on this proposal, we have implemented and evaluated in-depth a fully-fledged publish/subscribe broker running within Intel SGX, compiled in WebAssembly, and built on top of industry-battled frameworks and standards, \ie, MQTT and TLS protocols.
    Our extended TLS protocol preserves the privacy of attestation information, among other benefits.
    Our experimental results showcase most overheads, revealing a \benchmarkThroughputWasmSGXVsNative decrease in message throughput when using a trusted broker.
    We open-source the contributions of this work to the research community to facilitate experimental reproducibility. 
\end{abstract}

\def\confname{27th Conference on Principles of Distributed Systems (OPODIS'23)}
\def\confdoi{10.4230/LIPIcs.OPODIS.2023.23}

\definecolor{yellowPaper}{HTML}{fff8ae}
\AddToShipoutPictureFG*{\AtTextUpperLeft{\raisebox{66pt}{\newcommand{\copyrightmargin}{30pt}\newcommand{\leftmarginwidth}{\dimexpr(\oddsidemargin + 1in + \hoffset)}\hspace*{\dimexpr((\leftmarginwidth)*-1 + \copyrightmargin)}\begin{tcolorbox}[width=\dimexpr(\paperwidth-\copyrightmargin-\copyrightmargin),colback=yellowPaper,enhanced,frame hidden,sharp corners]  
            \centering\scriptsize
            This is the author's version of the work. The definitive version has been published  in the proceedings of the\\
            \confname.
            \href{https://doi.org/\confdoi}{DOI: \confdoi}
        \end{tcolorbox}   
    }}}%

\section{Introduction}\label{sec:intro}
Publish/subscribe (\pubsub) systems~\cite{DBLP:journals/csur/EugsterFGK03} have become foundational for seamless intercommunication of a wide range of devices, from IoT ecosystems to large-scale cloud-based services, \ie the cloud-edge continuum~\cite{DBLP:conf/icwe/TaivalsaariMP21}.
These systems enable efficient and scalable data distribution among distributed entities by decoupling data producers from data consumers.
Given their widespread adoption, notably in cloud computing~\cite{Amazon2023Pubsub,Google2023Pubsub,Microsoft2023Pubsub}, several \pubsub systems have been proposed with the clear goal of providing additional privacy guarantees~\cite{DBLP:journals/csur/OnicaFMR16}. 
However, the nodes participating in these systems are implicitly trusted, relegating security concerns primarily to the protection of communication channels or leveraging heavyweight cryptographic primitives~\cite{DBLP:conf/opodis/JoaquimP017}.
This limited security approach leaves data on the processing components vulnerable to potential threats, especially in decentralised and heterogeneous environments.
Notably, high-privileged actors within the \pubsub nodes, \ie, operating system or hypervisor, may compromise the confidentiality and integrity of data.
Similarly, they could leak critical cryptographic material, \eg, private keys of certificates.
Leaking certificate keys is especially concerning as these keys serve as the foundation for the authentication process among \pubsub participants, thereby putting user privacy and data integrity at stake.

In both the consumer market and cloud providers, trusted execution environments (TEEs) present a solution to strengthen the integrity and confidentiality of data in use, especially on nodes that may not be inherently trustworthy.
TEEs provide \emph{enclaves}, \eg, hardware-protected memory regions, where sensitive computations are completely isolated from other software executed on the same platform.
Such secure enclaves significantly elevate the security posture of systems like \pubsub, safeguarding not just the communication but also the processing and data on such nodes from malicious actors.
As a keystone feature of TEEs, attestation enables a remote entity to verify the authenticity, configuration, and state of a trusted environment using cryptographic proofs, ensuring the enclave runs the intended software without being tampered with or compromised~\cite{DBLP:conf/dais/MenetreyGKPFSR22}.
In \pubsub systems deployed over untrusted infrastructures, attestation ensures data and its processing within the TEE enclave are kept confidential and untampered.

Nonetheless, developing applications for TEEs is challenging due to their specific programming paradigms and SDKs, requiring massive efforts when writing or porting existing software~\cite{DBLP:conf/dais/GottelFS19}.
In addition, while many \pubsub systems exploit TEEs to protect data in use (covered in \S\ref{sec:rw-pubsub}), these usually tie closely to a specific TEE architecture, which is limiting in the heterogeneous environments (\eg using different CPUs) considered in this work.
Beyond this, current secure communication protocols fail to transport attestation proofs, do not maintain the privacy of these messages when supported, or cannot expose X.509 certificates issued by global authorities (covered in \S\ref{sec:rw-tls}).
This leads to the use of ad-hoc implementations, which fall short in offering a consistent solution across the cloud-edge continuum.

In this paper, we solve the major challenges associated to writing secure and portable \pubsub systems by relying on WebAssembly (\wasm), an open-standard binary instruction format.
\wasm's architecture-neutral design abstracts the complexity of hardware and TEE requirements, making it particularly suitable as a compilation target for \pubsub systems in heterogeneous hardware environments, such as the cloud-edge continuum.
We further protect the communication channels by extending the industry standard TLS protocol for embedding attestation evidence in the handshake while maintaining compatibility with the original specifications.
This ensures the authenticity of the executing code of parties within the \pubsub system, thereby mitigating the risk of malicious entities impersonating or modifying these components.
Additionally, certificate keys are safeguarded within the TEEs, preventing potential leaks.
Besides, we developed a proof-of-concept that encapsulates the full implementation of a standard \pubsub broker and a TLS library, enabling the termination of TLS channels directly in the TEE, which has been proven complex in prior work~\cite{talos2017aublin}.
We achieve this by using \wasm as a compilation target for both software, limiting the number of code changes required to make them compile and run within TEEs.
While our prototype is focused on cloud environments by leveraging Intel SGX, we outline the trusted primitives required for our proposal to be compatible with other platforms, including edge and IoT devices.
We summarise our contributions as follows:
\emph{(1)}~a unified strategy that secures a standard \pubsub system using TEEs with attestation for security, and compiled in \wasm for seamless cloud-edge communication while minimising code changes,
\emph{(2)}~an extension to the TLS communication protocol, facilitating the confidential exchange of attestation evidence, thereby affirming the authenticity of actors within the \pubsub system and
\emph{(3)}~an open-source implementation of a \pubsub broker using Intel SGX for cloud environments, with a suite of benchmarks aimed at evaluating the impacts of \wasm, the TEE and attestation, in comparison with state-of-the-art work.
Our evaluation reveals that our system delivers messages at \benchmarkThroughputWasmSGXVsNative slower than baseline throughput, yet provides portability and the robust security guarantees of TEEs.

\section{Background}\label{sec:background}
\subsection{\Pubsub systems}

A publish/subscribe system (often called \pubsub) is an asynchronous architecture for message passing that connects two different types of entities: publishers and subscribers.
Publishers send messages (or events), being unaware of who is interested in receiving such messages.
Messages sent are usually marked with an arbitrary type or a category.
Subscribers express their interest in specific types of messages and receive them when they are published.
\Pubsub is commonly used in event-driven distributed systems such as the Internet-of-things, since they simplify the communication between different entities.

\Pubsub messages are usually passed from publishers to subscribers using an intermediary broker.
Brokers receive messages from publishers and forward them to subscribers who have expressed interest in the corresponding topics.
Message brokers are responsible for routing and delivering messages to the appropriate subscribers.
Examples of message brokers include Apache Kafka, RabbitMQ, MQTT brokers, and cloud-based pub/sub services like Amazon SNS (Simple Notification Service) or Google Cloud Pub/Sub.

One of the key advantages of \pubsub is decoupling since publishers and subscribers do not need to know each other.
\Pubsub systems can also be designed to be highly scalable and fault-tolerant by using replicated, interconnected brokers.
This flexible architecture allows components to be added or removed without affecting the system.

Security in \pubsub systems is dealt with when publishers and subscribers connect to their brokers.
Brokers then usually implement authentication and access control before publishing or subscribing to messages.
They can also establish encrypted connections (\eg TLS) to implement message privacy.
By implementing attestation, we offer a stronger guarantee for all the components.
Publishers, subscribers, and brokers are guaranteed to be who they say they are, and all can ensure that their counterparts execute appropriate (correct) software.

\subsection{Trusted execution environments}
Trusted execution environments (TEEs) offer secure and isolated areas of a computer system.
TEEs shield sensitive data and code from unauthorised access, compromised libraries or operating systems.
This isolation is hardware-enforced, considerably minimising the software-based vulnerabilities.
In contrast to trusted platform modules (TPMs)~\cite{iso2015information}, which are limited to a specific and finite set of operations, TEEs enable the execution of arbitrary code, providing greater flexibility and versatility.
Code executed inside a TEE can communicate with untrusted programs or libraries, but such interactions are tightly controlled.
TEE-specific mechanisms ensure data is secure during transfers into and out of an enclave.

We focus on Intel Software Guard Extensions (SGX)~\cite{DBLP:journals/iacr/CostanD16,DBLP:journals/pomacs/NgocBBTSFH19}, a popular and widely used TEE architecture.
Available in server-grade CPUs, Intel SGX manages and runs code within secure enclaves, which interact with the other processes and operating system by transitioning the enclave boundary using specific instructions, \ie, \texttt{ECALLs} and \texttt{OCALLs}. 
Enclaves use \texttt{ECALLs} to receive function calls from untrusted applications outside the enclave and \texttt{OCALLs} to invoke functions outside from within the enclave.
The enclave's memory is encrypted by a dedicated MMU, ensuring protection against attackers with access to memory data.

\subsection{Attestation}
Attestation is a security mechanism enabling an \emph{attester} to prove its identity and integrity to an \emph{verifier}.
In the realm of TEEs, attestation allows a TEE to prove its configuration, identity, and state to another device or service.
Most TEE implementations support attestation, either built-in (as in SGX~\cite{johnson2016intel} and its successor TDX~\cite{DBLP:journals/access/SardarMF21,DBLP:journals/corr/abs-2303-15540}, AMD SEV-SNP~\cite{sev2020strengthening}) or by means of research prototypes~\cite{DBLP:conf/icdcs/MenetreyPFS22}.
The primary objective is to ensure the attester is genuine, unmodified, and trustworthy.
To achieve this objective, attestation leverages proofs (\ie \emph{evidence}), a cryptographically signed document composed of \emph{claims}, \ie, pieces of asserted information, like the hash of a code running in the TEE, \ie, \emph{measurement}.
The verifier is bootstrapped with \emph{reference values}, compared against the received claims for validation.

In Intel SGX, attestation is paramount for establishing the authenticity and integrity of code and data inside an enclave.
Intel controls a set of keys, intrinsically linked to the hardware and identity of the enclave.
Such keys are inaccessible from the enclaves themselves, and are used by the processor and Intel signed software to issue trustworthy evidence.
Intel is considered trusted, as they provision a subset of keys into the hardware. 
Within this assumption, third parties can trust evidence produced by the TEE.

\subsection{Communication channel and attestation binding}
\subparagraph*{Communication channel}
Over the past decades, several standards have emerged to establish trustworthy communication between two entities.
Among them, Transport Layer Security (TLS) stands out as the most prominent protocol for such tasks.
This cryptographic protocol ensures confidentiality, integrity, and authentication for secure communication.
A typical TLS session begins with a phase known as the \emph{handshake}.
This TLS handshake acts as a negotiation process, where the two parties decide on encryption settings and authenticate one another before exchanging secure data.
More precisely, the latest version of the TLS handshake (1.3 at the time of this writing) is composed of
\emph{1)} the \clienthello sent by the client, notably containing the supporting cipher suites, key agreement and anti-replay mechanisms,
\emph{2)} the server sends the \serverhello response, detailing analogous parameters, while signifying the handshake's conclusion, and finally
\emph{3)} the handshake wraps up as the client reciprocates with a similar acknowledgement, accompanied by data from a higher-level protocol, that is wrapped within TLS, \eg, HTTPS.
Additionally, during steps 2 and 3, both entities can opt to present an X.509 certificate.
These serve as identity validators, embedding trust through a chain of digital signatures rooted in trusted Certificate Authorities (CAs).
Building on this, many existing attestation protocols that bind with TLS typically augment the handshake, the certificate, or a combination of the two, facilitating the negotiation of security parameters and the exchange of attestation evidence.

\subparagraph*{Channel binding}
Channel binding~\cite{DBLP:conf/spw/AsokanNN03} ensures that an entity participating in a secure communication, typically via protocols such as TLS, is indeed the entity that has undergone attestation.
It establishes that the communications are with the attested environment, preventing possible relay attacks where a malicious party might relay attestation challenges to a genuine system and then claim the legitimate evidence as their own.
To secure trusted communication, attestation evidence is typically integrated into the early phases of the communication protocol.
However, combining attestation with these protocols introduces challenges, including increased latency and additional trade-offs like binding the system to a particular TEE technology.

As later outlined in this paper, the proposed solution refines the TLS protocol with minimal enhancements.
Unlike traditional approaches that embed attestation in custom fields in the broker's X.509 certificates, our method leverages a custom TLS 1.3 encrypted extension.
This approach reduces the need for additional communication roundtrips between the client and server.
It preserves the broker's ability to use certificates issued by recognised CAs, an aspect overlooked in previous studies.

\subsection{WebAssembly}
WebAssembly (\wasm)~\cite{DBLP:conf/pldi/HaasRSTHGWZB17} is an open standard that provides a portable and executable bytecode, tailored for running applications at near-native speed.
While its initial design targeted web applications written in traditional native languages like C/C++, Rust, and Go, its remarkable performance has led to its adoption not just within browsers but also in standalone environments~\cite{DBLP:conf/pldi/HaasRSTHGWZB17,DBLP:conf/iiswc/Wang22}.
\wasm was developed as a cross-platform, language-agnostic compilation target, optimised for easy integration in resource-limited environments, making it a prime candidate for TEE development.
Notably, several specialised runtimes, such as \textsc{Twine}~\cite{DBLP:conf/icde/MenetreyPFS21,Enarx} for Intel SGX and \textsc{WaTZ}~\cite{DBLP:conf/icdcs/MenetreyPFS22} for Arm TrustZone, have been developed to leverage \wasm's capabilities within state-of-the-art TEE implementations, and natively support key TEE primitives such as attestation.
For Wasm applications, particularly in standalone settings, the WebAssembly System Interface (WASI)~\cite{Mozilla2019StandardizingWASI} provides direct access to essential OS services, traditionally managed by POSIX and system calls.
WASI facilitates seamless communication with a handful of system components, ranging from file systems and networking to time management and random number generation, regardless of the underlying platform.
Runtimes optimised for TEEs map these system calls to trusted services, typically offered by the TEE manufacturers, ensuring that sensitive data remains secure and protected within the enclave.

We leverage \wasm as a versatile, cross-platform environment to model our \pubsub proposal seamlessly across the cloud-edge continuum and within TEEs as the execution context.
This approach offers unmatched portability across many processor and TEE architectures. 
When combined with TEEs, \wasm hardens our solution against powerful threats, such as the operating system, the hypervisor or malicious system administrators.
This security guarantee is especially important in edge computing scenarios where the local infrastructure typically lacks physical security measures as established in cloud environments.

\section{Related Work}\label{sec:rw}

\newcommand{\YES}{\compfull\xspace}
\newcommand{\HALF}{\comppart\xspace}
\newcommand{\NO}{\compnone\xspace}
\newcommand{\notapplicable}{---}
\newcommand{\notavailablesymbol}{\textsuperscript{\textdagger}}
\newcommand{\notavailable}{\compunknown\xspace}

\begin{table}[!t]
    \centering
    \footnotesize
    \setlength{\tabcolsep}{3pt}
    \rowcolors{1}{gray!0}{gray!10}
    \begin{tabularx}{\textwidth}{Xw{c}{5em}w{c}{5em}w{c}{5em}w{c}{5em}w{c}{5em}|w{c}{5.5em}}
        \toprule
        \rowcolor{gray!25}
        &\makecell{\scriptsize\cite{DBLP:conf/middleware/PiresPFF16}\\\scriptsize\textsc{SCBR}}&\makecell{\scriptsize\cite{DBLP:conf/srds/ArnautovBFFGKOM18}\\\scriptsize \textsc{PubSub-SGX}}&\makecell{\scriptsize\cite{DBLP:conf/trustcom/WangPF021}\\\scriptsize\textsc{MagikCube}}&\makecell{\cite{DBLP:conf/srds/SegarraDS20}\\\scriptsize\textsc{MQT-TZ}}&\makecell{\scriptsize\cite{DBLP:journals/cybersec/PeiSFSLYSM23}\\\scriptsize Pei\\[-1pt]\scriptsize\emph{et al.}}&\makecell{\scriptsize This\\[-3pt]\scriptsize work}\\
        \midrule
        Comm. protocol&TLS&TLS&TLS&TLS&\notavailable&TLS\\
        Fully enclaved broker&\YES&\YES&\YES&\NO&\NO&\YES\\
        Peer authentication&\NO&\YES&\YES&\YES&\notavailable&\YES\\
        Peer attestation&\NO&\NO&\NO&\NO&\notavailable&\YES\\
        Broker authentication&\YES&\YES&\YES&\YES&\notavailable&\YES\\
        Broker attestation&\HALF&\YES&\YES&\NO&\notavailable&\YES\\
        Persistence of messages&\NO&\YES&\NO&\NO&\NO&\YES\\
        Idiomatic \pubsub arch.&\NO&\YES&\YES&\YES&\YES&\YES\\
        Open source&\NO&\NO&\NO&\YES&\NO&\YES\\
        TEE technology\phantom{\YES}&SGX&SGX&SGX&TZ&SGX&\makecell{\textit{Agnostic}}\\
        \bottomrule
        \noalign{\vskip 2pt}
        \rowcolor{white}
        \multicolumn{7}{l}{\scriptsize TEEs references: SGX (Intel SGX), SNP (AMD SEV-SNP), TDX (Intel TDX), TZ (Arm TrustZone)}\\
    \end{tabularx}
    \caption{Comparison of the state-of-the-art \pubsub systems shielded by TEEs. \YES, \HALF, \NO mean fully, partially and not implemented, respectively. \notavailable denotes not disclosed details. Features description:
    \emph{1) Communication protocol}: protocol used by peers to interact with the broker.
    \emph{2) Fully enclaved broker}: broker operates within the TEE instead of only securing specific components within the secure environment.
    \emph{3) Peer authentication}: broker authenticates the peers while establishing communication.
    \emph{4) Peer attestation}: broker attests the peers while establishing communication.
    \emph{5) Broker authentication}: peers authenticate the broker while establishing communication.
    \emph{6) Broker attestation}: peers attest the broker while establishing communication; \HALF means that only publishers are attested.
    \emph{7) Persistence of messages}: system's capability to store messages for future delivery (\eg, when subscribers might be temporarily offline).
    \emph{8) Idiomatic \pubsub architecture}: system adheres to the principles of the \pubsub paradigm.
    \emph{9) Open source}: implemented solution is freely available to the public via an open-source repository.
    \emph{10) TEE technology}: denotes the TEE used by the proposed system, or its capacity to operate agnostically across various TEEs.}
    \label{table:rw-pubsub}
    \vspace{-10pt}
\end{table}

\subsection{\Pubsub with TEEs}\label{sec:rw-pubsub}
In the dynamic world of distributed systems, \pubsub mechanisms have consistently gained traction, acting as the core foundation for a variety of applications and architectures.
Many approaches have been conducted in research to bring dependability and safety regarding \pubsub systems, and in many different directions~\cite{wang2002security}.
An explored aspect is encryption schemes of data transferred through the brokers, preserving the privacy of the communicated information~\cite{DBLP:journals/cn/IonRC12,DBLP:conf/nss/NabeelABB13,DBLP:journals/tdsc/BarazzuttiFMOR17,DBLP:journals/fgcs/BorceaGPRR17,DBLP:conf/IEEEares/MalinaSDHF19,DBLP:conf/srds/GaballahCSM21,DBLP:conf/dbsec/BerlatoMCR22}.
Cryptographic-based privacy protection schemes focus on encrypting events and subscriptions, and then performing ciphertext matching between them. However, they often suffer from scalability issues as matching time complexity grows with the number of subscriptions, leading to diminishing performance.

More recently, researchers have investigated the potential of Intel SGX as a secure environment for confidential data processing.
Leveraging TEEs has shown potential for enhanced performance over cryptography-based methods, as highlighted by \textsc{SCBR}~\cite{DBLP:conf/middleware/PiresPFF16}.
Their study details a custom content-based routing engine operating within an SGX enclave.
\textsc{PubSub-SGX}~\cite{DBLP:conf/srds/ArnautovBFFGKOM18} introduced a scalable approach, using a load balancer that manages multiple matchers, each operating within individual enclaves.
Following this, \textsc{MagikCube}~\cite{DBLP:conf/trustcom/WangPF021} added an authentication service to the \pubsub system, thereby enhancing broker trust among publishers and subscribers using SGX.
Finally, Pei \emph{et al.}~\cite{DBLP:journals/cybersec/PeiSFSLYSM23} further refined this paradigm by optimising subscription matching times using cryptographic methods, with SGX facilitating comparison tasks but does not disclose how secrets are exchanged.
In contrast with previous work, our proposal prioritises establishing an initial trust across all \pubsub participants by mutual attestation.
We do it by encapsulating conventional \pubsub systems within TEEs, ensuring genuine execution environments and trustworthy implementations.

Other prior studies have chosen a different strategy using TrustZone, Arm's TEE.
In this setup, the device is divided into the \emph{normal world} (the standard OS) and the \emph{trusted world} (the TEE).
\textsc{MQT-TZ}~\cite{DBLP:conf/srds/SegarraDS20} migrated the broker's data management component within this trusted world.
Publishers and subscribers negotiate symmetric keys with the broker, which are generated inside the TEE.
This approach ensures that data is encrypted during transmission.
Conversely, our proposal enhances the threat model of \textsc{MQT-TZ} by relying on entity attestation in the \pubsub system and hosting the TLS endpoint directly within the TEE, avoiding handling cryptographic materials outside the TLS protocol.

\subparagraph{Comparison}
\cref{table:rw-pubsub} offers a comprehensive summary of state-of-the-art proposals for securing \pubsub systems with TEEs, focusing on features relevant to our study.

\begin{table}[!b]
    \centering
    \footnotesize
    \setlength{\tabcolsep}{3pt}
    \rowcolors{1}{gray!0}{gray!10}
    \begin{tabularx}{\textwidth}{Xw{c}{4em}w{c}{4em}w{c}{4em}w{c}{4em}w{c}{4em}w{c}{4em}|w{c}{5.5em}}
        \toprule
        \rowcolor{gray!25}
        &\makecell{\scriptsize\cite{intelraendtoend}\\\scriptsize SGX\\[-1pt]\scriptsize EPID}&\makecell{\scriptsize\cite{DBLP:conf/IEEEares/ShepherdAM17}\\\scriptsize Shepherd\\[-1pt]\scriptsize\emph{et al.}}&\makecell{\scriptsize\cite{DBLP:conf/icdcs/MenetreyPFS22}\\\scriptsize\textsc{WaTZ}}&\makecell{\scriptsize\cite{DBLP:journals/corr/abs-1801-05863}\\\scriptsize RA-TLS}&\makecell{\cite{DBLP:conf/dsn/GregorOVPQAMSFF20}\\\scriptsize\palaemon}&\makecell{\scriptsize\cite{DBLP:conf/nordsec/NiemiPE21}\\\scriptsize TSL}&\makecell{\scriptsize This\\[-3pt]\scriptsize work}\\
        \midrule
        Baseline protocol&Custom&Custom&Custom&TLS&TLS&TLS&TLS\\
        No change in TLS spec.&\notapplicable&\notapplicable&\notapplicable&\YES&\YES&\YES&\YES\\
        Attestation privacy&\NO&\YES&\NO&\NO&\YES&\YES&\YES\\
        Mutual attestation&\NO&\YES&\NO&\YES&\NO&\YES&\YES\\
        Evidence per session&\YES&\YES&\YES&\NO&\NO&\YES&\YES\\
        Endpoint in enclave&\YES&\YES&\YES&\YES&\YES&\NO&\YES\\
        Attestation privacy&\NO&\HALF&\NO&\HALF&\notavailable&\HALF&\YES\\
        TEE-agnostic&\NO&\HALF&\HALF&\NO&\NO&\HALF&\YES\\
        Support global CAs&\notapplicable&\notapplicable&\notapplicable&\NO&\NO&\NO&\YES\\
        Open-source&\YES&\NO&\YES&\YES&\NO&\NO&\YES\\
        TEE technology\phantom{\YES}&SGX&TZ&TZ&SGX&SGX&\textit{Agnostic}&\textit{Agnostic}\\
        \bottomrule
        \noalign{\vskip 2pt}
        \rowcolor{white}
        \multicolumn{8}{l}{\scriptsize TEEs references: SGX (Intel SGX), SNP (AMD SEV-SNP), TDX (Intel TDX), TZ (Arm TrustZone)}
    \end{tabularx}
    \caption{Comparison of the state-of-the-art channel binding solutions. \YES, \NO mean fully and not implemented, respectively.  \notavailable denotes not disclosed details. "\notapplicable" denotes not applicable comparisons. Features description:
    \emph{1) Baseline protocol}: protocol upon which the proposal is built.
    \emph{2) No change in TLS specification}: proposal respects the TLS specifications, ensuring compatibility with pre-existing TLS implementations.
    \emph{3) Attestation privacy}: protocol maintains confidentiality of attestation evidence.
    \emph{4) Mutual attestation}: communicating entities engage in a mutual attestation process.
    \emph{5) Evidence per session}: each communication session is uniquely associated with specific attestation evidence.
    \emph{6) Attestation Privacy}: all attestation-related messages are encrypted; \HALF means that only a portion of the messages remains confidential.
    \emph{7) TEE-agnostic}: independent of any specific programming language or TEE-specific SDK; \HALF means theoretical approach proposed, but no agnostic implementation.
    \emph{8) Endpoint in enclave}: endpoint application fully resides within the TEE.
    \emph{9) Support global CAs}: broker-displayed certificates can be vouched for by globally recognised CAs.
    \emph{10) Open source}: implemented solution is freely available to the public via an open-source repository.
    \emph{11) TEE technology}: denotes the TEE that the attestation API currently supports, or its capacity to operate agnostically across various TEEs.}
    \label{table:rw-tls}
    \vspace{-10pt}
\end{table}

\subsection{Binding Communication Protocols and Attestation using TEEs}\label{sec:rw-tls}
Research has extensively explored the integration of secure communication protocols with attestation.
While our focus primarily lies on solutions leveraging TEEs, we also recognise the significant contributions from prior work that leveraged TPMs as trust anchors~\cite{DBLP:conf/ccs/GoldmanPS06,stumpf2006robust,DBLP:conf/ccs/GasmiSSUA07,DBLP:conf/ccs/StumpfFKE08,DBLP:conf/ccs/ArmknechtGSSURV08,DBLP:conf/trustcom/YuWLY13,DBLP:journals/jcm/AzizUM14,paverd2015a,DBLP:conf/ucc/WalshM17,DBLP:conf/stm/WagnerBB20}.
Readers can refer to \cite{DBLP:journals/corr/abs-1801-05863,DBLP:conf/nordsec/NiemiPE21} for a comprehensive review of these works.
Our analysis omits explicitly works that discuss attestation through TEEs but do not bind attestation evidence to a communication channel~\cite{DBLP:conf/sp/McCuneLQZDGP10,DBLP:conf/uss/CostanLD16,DBLP:conf/sosp/FerraiuoloBHP17,DBLP:conf/ndss/WeiserWBMMS19}.

Intel proposed a remote attestation protocol for key exchange based on \textsc{SIGMA}~\cite{DBLP:conf/crypto/Krawczyk03,intelraendtoend}, binding SGX enclaves with communication channels.
Subsequent solutions aimed to establish trusted communication channels with enclaves using custom message exchanges~\cite{DBLP:conf/IEEEares/ShepherdAM17,DBLP:conf/eurosys/LeeKSAS20,DBLP:conf/icdcs/MenetreyPFS22}.
While these initial efforts in remote attestation provided valuable insights, their custom nature makes them challenging to integrate into existing software.
In contrast, our approach harnesses TLS, the leading industry standard for secure communication.

More recently, further research~\cite{TCG2021DiceAttestation, DBLP:conf/dsn/GregorOVPQAMSFF20,DBLP:conf/nordsec/NiemiPE21,DBLP:journals/corr/abs-1801-05863,DBLP:conf/fruct/NiemiSE22} have suggested using TLS for communication and modifying the X.509 certificates in order to include additional fields related to attestation.
Although this method takes advantage of the protocol's standardisation to address the earlier concern, it ties the attestation mechanism directly to the certificates exposed by the endpoints.
This direct connection implies that certificates must be dynamically generated, which restricts them from being signed by global CAs like \emph{Let's Encrypt}, commonly used for domain validation certificates.
Opting for a different route, we used the encrypted extensions of the TLS protocol for carrying evidence of the server.
Consequently, our approach is compatible with certificates endorsed by global CAs.
This is particularly beneficial if the endpoint owns a separate network-level identity, like a DNS or domain name.

\subparagraph{Comparison}
\cref{table:rw-tls} offers a comprehensive summary of cutting-edge research dedicated to binding communication channels with attestation, focusing on features relevant to our study.

\section{Design Considerations}\label{sec:design-considerations}
We explain the threat model of our design, highlight its security goals and cover the trusted primitives that must be available on a system to support our proposal.

\subsection{Threat Model}
Our approach relies on a few key trusted components, which are essential for our system to be deemed trustworthy.
We further discuss these elements in the remainder of this section.

\subparagraph{TEEs}
Our proposal leverages the protection offered by TEEs for securing the execution of applications and enforcing strong isolation against powerful attackers, such as the OS or the hypervisor.
Given that our proposal is TEE-agnostic, we highlight the minimal requirements to uphold trust in the \pubsub system and attestation mechanism.
We assume the application code can be inspected but cannot be subverted.
Data in use remains confidential and cannot be read unless granted by the TEE.
The hardware and software strictly required to run a TEE instance are considered trusted.
Furthermore, the TEE offers a remote attestation mechanism backed by genuine entities responsible for validating the trustworthiness of attestation evidence.
Although we do not address side-channel or denial-of-service attacks~\cite{DBLP:conf/wistp/BukasaLBLL17,DBLP:conf/ccs/WernerMAPM19,DBLP:journals/csur/FeiYDX21}, there exist measures for these in various TEE designs~\cite{DBLP:journals/corr/abs-2308-06442}.

\subparagraph{OS}
The OS follows an \emph{honest-but-curious} threat model, posing no threat to the trusted environment but interested in gathering sensitive information.
Consequently, it can monitor all communication within the \pubsub system.
Assuming the OS starts to behave maliciously, the trusted computing base (TCB) remains confidential and doesn't malfunction, though it might become unresponsive.
The applications running in the TEE are carefully developed to ignore abnormal responses and abandon execution in such cases.

\subparagraph{Wasm}
We presume that the \wasm runtime is implemented correctly and does not contain vulnerabilities.
The \wasm runtime acts as a shim library by encapsulating \wasm applications and uses trusted APIs from the TEE SDK for system interactions or sanitises the interaction with the untrusted OS when no secure option is available.
While side-channel attacks might target \wasm and TEEs~\cite{DBLP:journals/corr/abs-2212-07899}, they fall beyond the scope of this study.

\subparagraph{Cryptography}
As we embed a TLS library and enhance it to integrate attestation concerns, we suppose the correct implementation of the cryptographic ciphers and operations.
Additionally, we also presume that standard cryptographic techniques cannot be subverted and are hardened against side-channel attacks.

\subsection{Security requirements}
We propose a series of requirements for establishing trusted communication channels between the actors of \pubsub systems:\\

\noindent
\begin{tabularx}{\textwidth}{@{}l@{\hspace{2mm}}X}
    \textbf{(SR1):} & \textbf{Support global CAs certificates}: brokers shall support exposing certificates issued by globally recognised CAs.\\
    \textbf{(SR2):} & \textbf{Trust assurance}: \pubsub actors shall attest the TCB of the participant they connect with and must be similarly verified in return.\\
    \textbf{(SR3):} & \textbf{Channel and attestation bindings}: communication channels must be linked to newly created attestation evidence, preventing replay or collusion attacks~\cite{DBLP:conf/nordsec/NiemiPE21}.\\
    \textbf{(SR4):} & \textbf{Attestation privacy}: attestation information shall remain confidential between the endpoints of a given communication channel.\\
    \textbf{(SR5):} & \textbf{\Pubsub privacy}: all the data bound to the \pubsub system shall be inaccessible to outside actors, including the OS, kernel and hypervisor of the broker and peers.\\
    \textbf{(SR6):} & \textbf{\Pubsub narrow scope}: the peers shall only publish or subscribe to the topics as required for their needs.\\
\end{tabularx}

\subsection{Trusted primitives}\label{sec:trusted-primitives}
We identified a core set of trusted primitives that any system must support to host our proposal securely.
These primitives are provided by the secure environment or the \wasm runtime:

\begin{itemize}
    \item \textbf{Isolated execution context}: this ensures the runtime remains secure and unaffected by any other applications running concurrently on the system.
    All major TEE manufacturers support at least one \wasm runtime, though the isolation paradigm and threat model vary.
    \item \textbf{Attestation capabilities}: the TEE must expose two primitives to the \wasm runtimes: \emph{generate} and \emph{verify}.
    The former primitive generates evidence for proving the trustworthiness of the secure environment with additional data attached, such as a nonce and a public key, while the latter confirms the validity of this proof.
    \item \textbf{Network communication}: the \wasm runtimes require access to a network API that handles socket operations and the transfer of data.
    \item \textbf{File system access}: if the \pubsub broker needs to store publications for future delivery, the system should offer secure ways to save and access these files.
\end{itemize}

Other concerns, such as encryption and \pubsub protocol logic, are platform-independent and addressed using dedicated software compiled into \wasm.

\section{Architecture}
In this section, we outline the architecture and design rationale of our proposal.
First, we explain the changes made to the TLS protocol.
Next, we discuss the security enhancements of our \pubsub system.
We wrap up with insights into specific implementation details.

\subsection{Overview}

\begin{figure}
    \centering
    \includegraphics[height=5cm]{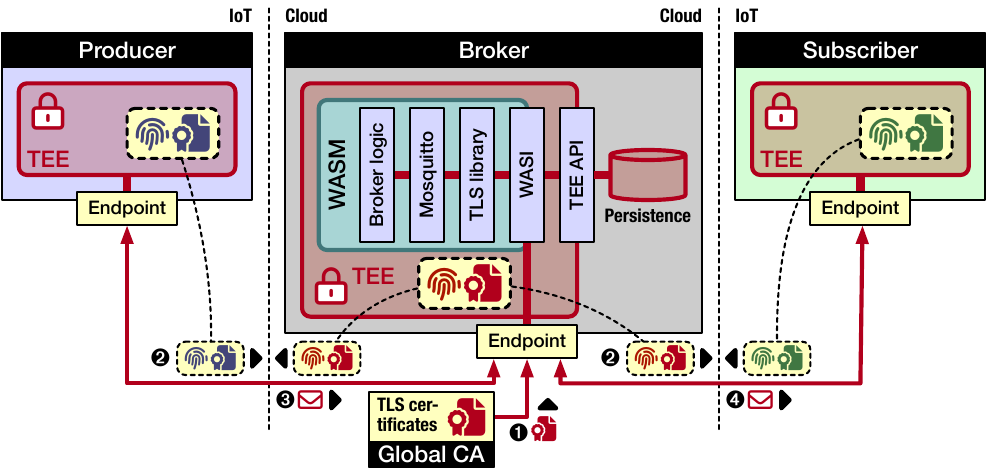}
    \caption{The overall architecture of our proposal. (\includegraphics[height=7px]{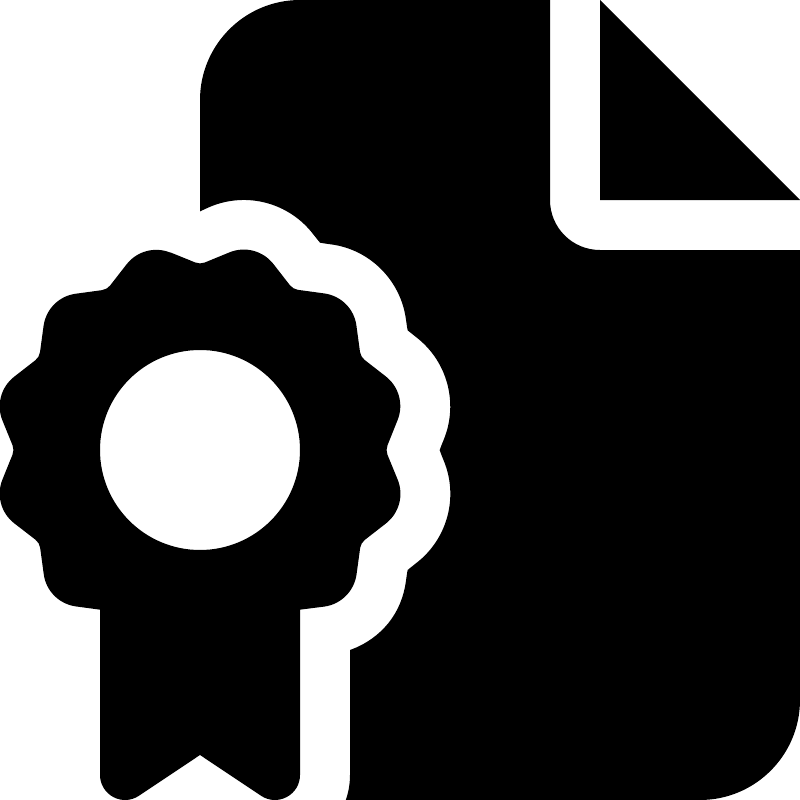}, \includegraphics[height=7px]{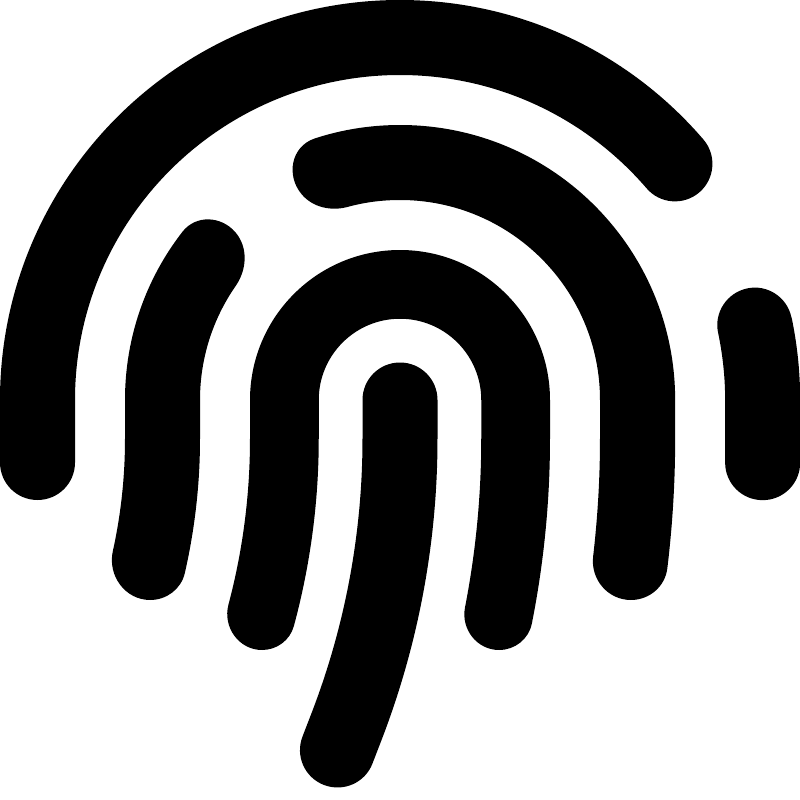}) mean X.509 certificate and attestation evidence, respectively. The colours of these icons correspond to the actor owning them.}
    \label{fig:architecture}
\end{figure}

We designed our proposal as a versatile system capable of running on numerous processor architectures.
The system isolates security-sensitive \pubsub operations of peers and brokers, ensuring the authenticity of connecting machines using TEEs and mutual attestation.
A key aspect of our design is the adoption of \wasm, facilitating cross-platform support across various TEE architectures.
More specifically, we rely on trusted \wasm runtimes~\cite{DBLP:conf/icde/MenetreyPFS21,DBLP:conf/icdcs/MenetreyPFS22}, to host a secure \pubsub system with its associated dependencies such as TLS libraries, thus covering the large spectrum of the cloud-edge continuum.

\Cref{fig:architecture} illustrates the key components and entities within our \pubsub solution.
Serving as the central hub, the broker first acquires a certificate from a global CA for its TLS endpoint (\DingBlack{1}).
Peers then initiate secure communication with the broker via our enhanced TLS handshake, performing mutual remote attestation by exchanging their X.509 certificate for authentication and attestation evidence (detailed in \cref{sec:attestation}).
This ensures that the broker and peers are trustworthy (\DingBlack{2}).
The publisher generates data and transmits it to the broker's TEE (\DingBlack{3}).
The broker, in turn, relays this data to the subscriber's TEE (\DingBlack{4}).
Both the publisher and the subscriber use the same technology stack in their respective TEEs, although this detail is omitted in the figure for clarity.

\subsection{Attesting communication channels}
\label{sec:attestation}

\begin{figure}
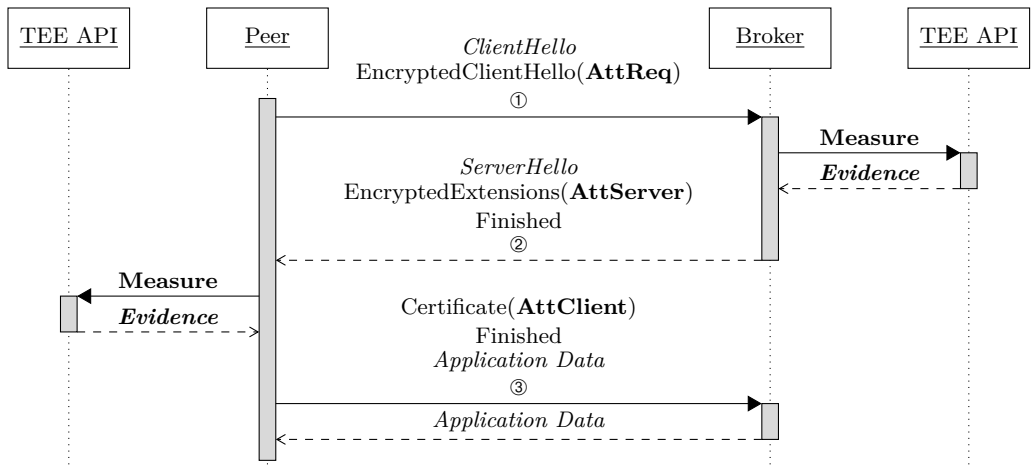

    \centering
    \begin{sequencediagram}
        \footnotesize
        \renewcommand\unitfactor{0.475}  \newinst{peertee}{TEE API}
        \newthreadshift[1]{peer}{Peer}
        \newinst[5]{broker}{Broker}
        \newinst[1]{brokertee}{TEE API}

        \begin{call}{peer}{\shortstack{\clienthello\\EncryptedClientHello(\attreq)\\\Ding{1}}}{broker}{\shortstack{\serverhello\\EncryptedExtensions(\attserver)\\Finished\\\Ding{2}}}
            \begin{call}{broker}{\textbf{Measure}}{brokertee}{\textbf{\textit{Evidence}}}
            \end{call}
            \postlevel \end{call}

        \begin{call}{peer}{\textbf{Measure}}{peertee}{\textbf{\textit{Evidence}}}
        \end{call}

        \postlevel 

        \begin{call}{peer}{\shortstack{Certificate(\attclient)\\Finished\\\textit{Application Data}\\\Ding{3}}}{broker}{\textit{Application Data}}
        \end{call}
        
    \end{sequencediagram}
    \caption{Enhanced TLS 1.3 handshake with attestation. New elements are mentioned in bold.}
    \label{fig:handshake}
\end{figure}

We enhanced the TLS protocol to integrate the exchange of attestation evidence when a peer is communicating with the broker.
This exchange of information occurs in the TLS handshake, as depicted in \Cref{fig:handshake}.
Our enhancements are highlighted in bold.

\subparagraph{Attestation protocol}
The first message of the handshake (\Ding{1}) is sent by the peer, which is comprised of an encrypted attestation request (\attreq), indicating that the peer is establishing a TLS channel requiring attestation.
Contrary to prior work, our protocol does not specify TEE architectures in this message, as our solution supports the verification of all the types of TEEs that can be used in the broker.
Similarly, our protocol is the only proposal that studies the encryption of the attestation request, as further explicated below.

When received, the broker answers with a message (\Ding{2}) composed of its evidence (\attserver), freshly generated and bound to the TLS session.
It is worth noting that we deliberately chose not to embed attestation data within the broker's X.509 certificate so the endpoint can use certificates endorsed by globally recognised CAs, satisfying \textbf{(SR1)}.

Subsequently, the peer verifies whether the evidence of the broker is genuine and checks if the evidence is bound to the current TLS session (to counter reuse attacks).
Moreover, it compares the code measurement of the broker to a known reference value, indicating that the code of the broker is trusted.
Since a global CA issues the broker's X.509 certificate, the peer verifies that the DNS or domain name of the broker matches the identifier exposed in the X.509 certificate~\cite{rfc6125}.
If these conditions are fulfilled, the peer issues its evidence (\attclient) and sends it to the broker (\Ding{3}), also bound to the TLS session and embedded within the peer's X.509 certificate.
We had to rely on a custom X.509 extension for the peer, as TLS 1.3 does not provide an extension point for the last message of the handshake.

Lastly, the broker also verifies whether the evidence of the peer is genuine, and checks if the evidence is bound to the current TLS session.
Similarly, the broker ensures that the code measurement of the peer matches a known reference value, indicating that the code of the peer is trusted.
Once completed, the peer and the broker mutually attest themselves, ensuring their respective TEE is trustworthy, satisfying \textbf{(SR2)}.
As a result, the TLS handshake is ended, and the applications may start \pubsub communication.

\subparagraph{Binding the handshake with attestation evidence}
Integrating the attestation mechanism in the TLS handshake has many benefits: the execution time is optimised since no additional round-trips are necessary, and the TLS 1.3 standard is respected by using extension points as designed by the protocol.
Besides, we strongly bind freshly generated evidence to individual TLS sessions, using unique TLS keying materials computable by both parties (RFC 5705)~\cite{rfc5705}, satisfying \textbf{(SR3)}.
This prevents replay and collusion attacks, which would affect published past evidence otherwise.
Since our \pubsub system and TLS library are compiled in \wasm, evidence binding within TLS handshakes is portable across different TEE architectures.

\subparagraph{Securing the attestation information}
We use three distinct encryption mechanisms to ensure the confidentiality of the attestation information, to comply with \textbf{(SR4)}.
First, we opted to use the recently-introduced \emph{encrypted client hello} (ECH) for TLS~\cite{ietf-tls-esni-16}, which is currently an IETF draft to communicate early information while preserving its privacy in the \clienthello message.
To the best of our knowledge, we are the first to leverage this draft to protect the attestation request (\attreq) in the TLS handshake.
In a nutshell, the \clienthello message is split into two parts: the outer message, which is in plain text, and the inner message, which is encrypted using a public key, typically distributed using DNS infrastructure.
Second, we rely on the TLS 1.3 encrypted extension in the broker reply (\ie, the \texttt{ServerHello} message), so the broker's attestation evidence remains confidential.
This has been made possible because the two parties can derive a shared secret for encryption up to this point.
Finally, the peer's certificate, which contains the peer's evidence, is also protected, since the entire third message of the handhake is encrypted by design.

\subsection{Securing \pubsub systems}
We leverage TEEs and trusted \wasm runtimes for implementing our \pubsub design, ensuring strong hardware isolation of both code and data.
This design remains versatile, working with various TEE architectures as long as they offer the trusted primitives for the \wasm runtimes (as in \S\ref{sec:trusted-primitives}).
When paired with the mutually attested TLS protocol, we shield data in use and in transit, satisfying \textbf{(SR5)}.
As a \pubsub software, we selected Mosquitto~\cite{DBLP:journals/jossw/Light17}, a well-known and open-source message broker that implements the latest version of the MQTT protocol.
We have chosen Mosquitto because it is lightweight and suitable for use on all devices, from low-power IoT devices to cloud servers.
Besides, we used WolfSSL, an embeddable cryptographic library, enabling the host of the TLS termination directly in the TEE, so external systems cannot eavesdrop on the communication, nor alter the code that maintains the endpoint.
We needed to compile Mosquitto in \wasm, which required slight modifications to the source code.
Regarding WolfSSL, we reused one of the extension works of a trusted runtime~\cite{DBLP:conf/icde/MenetreyPFS21}, which already compiled this library in \wasm.

\subparagraph{System interactions}
Mosquitto, like most software, is required to perform system calls for interactions with the outside world.
Typically, this is the case for the socket API when exposing the TLS endpoint.
For such usages, we rely on WASI, which translates the system calls to the underlying OS seamlessly.
As the private and session keys of TLS are located within the TEE, the communication remains confidential against eavesdropping attempts, even when system calls are monitored.
Besides, Mosquitto can persist undelivered messages in a database for later transmission to offline subscribers.
The WASI specification includes a file system API for data storage, while TEEs typically provide means to save files via a trusted API securely.
In contrast to the socket API, \wasm runtimes use that TEE API to transparently encrypt files, ensuring the confidentiality of the stored messages.

\subparagraph{Securing and narrowing the \pubsub data access}
Our proposal ensures that brokers and peers are trustworthy as they are mutually attested.
As such, we inherently restrict adversaries to propagate and observe messages going through the \pubsub system.
Nonetheless, we propose to reduce further the scope of peers using access control lists (ACLs), which is a built-in functionality of Mosquitto.
When paired with the X.509 certificates presented in the TLS handshake, this feature enables to authenticate peers without requiring additional usernames or passwords, as this is typically the case using Mosquitto.
This narrowed publication and subscription scope satisfies \textbf{(SR6)}.
Future work may further adapt the ACLs to be bound to evidence, provided that code measurements differ from each actor.

\subparagraph{Use cases}
Our approach is applicable to a variety of distributed \pubsub systems, which supports scalable communication by decoupling publishers from subscribers and ensures secure messaging through mutual attestation.
This was demonstrated in our European-funded project in collaboration with Siemens for secure data processing at edge nodes and capturing results on an MQTT messaging bus, as well as with the Byzantine fault-tolerant (BFT) system of the University of Lisbon for maintaining and ensuring trust in the attestation reference values~\cite{DBLP:conf/sp/VassantlalAFB22}.
To balance the need for compactness with computational capability, we used Intel NUCs as edge nodes, which are small, efficient PCs equipped with processors that support Intel SGX.
This allows the publisher on the edge nodes to verify whether the broker is genuine based on evidence and the code hash contained in that payload.
This code hash is then compared to the trusted reference values stored in the BFT-resilient system.
The evaluation section (\S\ref{sec:evaluation}) details how the broker performs when handling heavy loads from multiple publishers and subscribers, as encountered in these distributed environments.

\subsection{Implementation}
We developed a prototype of a trusted \pubsub broker using Intel SGX to better understand the practical impacts and performance overhead associated with our proposal.
As such, we used \textsc{Twine}~\cite{DBLP:conf/icde/MenetreyPFS21}, a trusted \wasm runtime designed to secure applications on Intel SGX.
We target the latest Intel SGX technology, named \emph{Intel Scalable}, for creating enclaves with an enclave page cache (EPC) up to \qty{512}{\gibi\byte}, whereas prior work was limited to \qty{128}{\mebi\byte}, which was a performance bottleneck when reaching that threshold.

\subparagraph{Generating attestation evidence}
We provided the two trusted primitives of attestation (as in \S\ref{sec:trusted-primitives}), \ie \emph{generate} and \emph{verify}, which use the Intel SGX attestation capabilities.
The former primitive calls the underlying TEE API to create a payload in the JSON format, which is later communicated in the TLS handshake.
That payload notably contains the type of TEE and the evidence itself.
The latter primitive receives a JSON and indicates whether the remote system is genuine.
The runtime abstracts these primitives using \emph{librats}~\cite{librats}, a library capable of generating and verifying attestation on Intel SGX, TDX and AMD SEV-SNP.
For each attester type, a matching verifier type is implemented in librats, which avoids introducing TEE architecture-specific concerns in the \wasm applications.
An enclave can produce a cryptographic summary of its state through the \texttt{EREPORT} instruction.
This summary, or report, includes assertions to ensure the TCB is updated and secure.
In addition, it also contains code measurements of the runtime and hosted \wasm application, along with the keying materials of the current TLS session.
The report is then forwarded to an Intel component known as the \emph{quoting enclave}, which fetches a report key using the \texttt{EGETKEY} instruction to authenticate the integrity of the report.
Upon successful verification, the quoting enclave generates evidence, termed a \emph{quote} in the context of SGX, allowing the other actor (\eg the publisher, subscriber or broker) to attest the enclave~\cite{DBLP:conf/dais/MenetreyGKPFSR22,DBLP:conf/icfem/SardarFF20}.
As a remote attestation mechanism, we use the Intel data center attestation primitives (DCAP)~\cite{scarlata2018DCAP}, which avoids involving Intel during the verification of the attestation evidence, speeding up the process and allows our \pubsub system to operate without an Internet connection.

\pagebreak \subparagraph{\Wasm}
We compiled Mosquitto using Clang with WASI-SDK~\cite{WasiSDK}, a toolchain that compiles C/C++ source code into \wasm for non-web environments.
Mosquitto lacks support for a portable and \wasm-enabled TLS library.
Therefore, we modified Mosquitto to use WolfSSL as a drop-in cryptographic library replacement.
We leverage the compatibility layer of WolfSSL with OpenSSL, minimising henceforth the required code changes.
Further, we disabled some of Mosquitto's system interaction layers (\eg, signals, dynamic library loading) as not supported by WASI.
Turning off these features allowed a smooth embedding of Mosquitto for \wasm into the SGX enclave.
The adaptation of Mosquitto required changes to 610 SLOC (in C source/header files), which accounts for 1.2\% of the total codebase.

\subparagraph{Securing the file system}
Our proposal relies on the trusted file system of \textsc{Twine}, which is backed by the Intel protected file system (IPFS).
IPFS uses AES-GCM for encryption, taking advantage of hardware acceleration from the CPU. 
Files are encrypted into a Merkle tree structure of \qty{4}{\kibi\byte} nodes and stored on the untrusted file system, with each node securing its children through encryption keys and tags.
When the enclave requests data, IPFS decrypts tree nodes within the shielded memory of the TEE, ensuring the information stored on the untrusted file system remains confidential and unaltered.
In our proof of concept, IPFS automatically generates the keys based on enclave signatures and processor-specific keys.
For more sophisticated use cases, such as fault tolerance, \wasm programs can provide their encryption keys, which can be obtained from an external key management system.

\subparagraph{Supporting the cloud-edge continuum}
We opted for Intel SGX for hosting our proof of concept because \textsc{Twine} supports all the trusted primitives outlined in Section \ref{sec:trusted-primitives}, which are necessary for running our proposal.
While alternative implementations are possible on other TEE architectures like Intel TDX and AMD SEV-SNP using the runtime WAMR~\cite{WAMR}, or Arm TrustZone using \textsc{WaTZ}~\cite{DBLP:conf/icdcs/MenetreyPFS22}, these options would require additional effort because they currently support only some of the trusted primitives we rely on.
For example, \textsc{WaTZ} enhances Arm TrustZone with attestation capabilities but does not offer file persistence or a complete socket API.
In the case of WAMR, its mechanisms for attestation have yet to be tested on AMD platforms. 
Nonetheless, we are confident that with some extensions to these runtimes, our prototype can be adapted to work with a variety of TEE architectures, enabling an agnostic approach for trustworthy \pubsub systems.

\section{Evaluation}\label{sec:evaluation}

Our evaluations assess several aspects of our proof of concept.
We intend to answer the following questions:
\emph{1)} What performance costs are associated with the use of \wasm, SGX, and the creation of attestation evidence on the broker during the connection process?
\emph{2)} How does our solution scale when increasing message throughput?
\emph{3)} How does our solution scale with a growing number of publishers?
To answer these questions, we measure the connection times with different peers (\S\ref{sec:eval-handshake}), stress the broker with a high volume of messages (\S\ref{sec:eval-messages}), and grow the number of publishers while observing the resulting latency (\S\ref{sec:eval-publishers}).

\subparagraph{Scope}
Our evaluation is deliberately centred on the broker component, which is the core and critical part of distributed \pubsub architectures.
As the broker orchestrates the message flow among publishers and subscribers, it largely influences the scalability of the system.
Our analysis examines the broker in different variants: native execution without SGX (the baseline), \wasm outside SGX, and \wasmSGX, running within SGX.
To ensure a consistent comparison, publishers and subscribers are executed in their native environment without SGX.
While our focus remains on the broker to evaluate performance benchmarks, a fully secure system requires that all entities operate within TEEs to protect the confidentiality of data, including publishers and subscribers.

\subparagraph{Experimental setup}
The broker runs on a 16-core Intel Xeon Gold 6326 (\qty{2.9}{\giga\hertz}), running Ubuntu 20.04, SGX2 (EPC of \qty{64}{\gibi\byte}) with SGX driver v1.41 (DCAP) and SDK v2.20.
The publishers and subscribers are executed on an 8-core Intel Xeon E3-1270 v6 (\qty{3.8}{\giga\hertz}), running Ubuntu 20.04.
These two machines are connected through a \qty{1}{\giga\bit\per\second} switch.
We modified Mosquitto v2.0.15 and WolfSSL v5.5.3, and compiled them using Clang 10 at maximum optimisation.
For \wasm benchmarks, we initially compiled the application into \wasm bytecode with Clang, and used ahead-of-time compilation into assembly using WAMR's compiler at level size 1, as required for SGX.
Time is measured using Linux's monotonic clock, including enclave exit and re-entry times (\benchmarkRoundtripWasmSGX).
The MQTT broker is configured to match the minimum QoS offered, \ie fire-and-forget mode.\footnote{MQTT supports two other operating modes, respectively delivering the message at least once with confirmation required and delivering it exactly once. Our proposal avoids the overheads of a four-step handshake while guaranteeing both the integrity and authenticity of the involved peers.}
Our implementation is open-source, and instructions to reproduce our experiments are available on GitHub~\cite{Menetrey2023Opodis}.

\subsection{Establishing new connections}
\label{sec:eval-handshake}

\begin{figure}[!t]
    \centering
    \includegraphics[width=\textwidth]{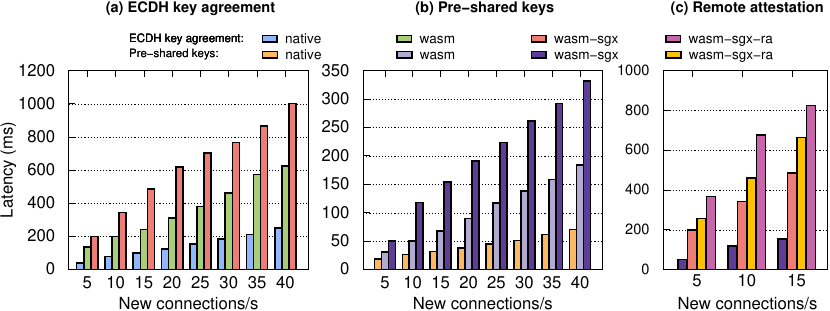}
    \caption{Latency for each new connection at varying connections per second.}
    \label{fig:case-1}
\end{figure}

Our initial evaluation measures the average latency incurred when a varying number of peers attempt to establish new connections with a broker.
In this setup, we instruct a certain number of peers (depicted on the x-axis) to start a new connection every second.
The y-axis displays the resulting latency.
\Cref{fig:case-1}a depicts the latency for connections made using TLS handshakes with Elliptic-curve Diffie-Hellman (ECDH) as a key agreement protocol.
\Cref{fig:case-1}b shows the latency when using a pre-shared key (PSK), effectively bypassing asymmetric cryptography.
The latter approach is applicable when a peer has previously established a connection with the broker and already has a mutually agreed-upon secret key.

Across both (ECDH and PSK) scenarios, latency scales linearly for every variant as the number of peers connecting per second increases from 5 to 40.
We observe higher latencies with \wasm (using ECDH) over native execution with an overhead of \benchmarkNewConnectionsWasmVsNative, due to \wasm's performance constraints relative to native execution~\cite{DBLP:conf/usenix/JangdaPBG19}.
Additionally, \wasmSGX (using ECDH) exhibits further overhead (\benchmarkNewConnectionsWasmSGXVsNative) compared to native because of the enclave, which is expected given the added security benefits.
A contributing source of this added overhead is the need for frequent switching between the enclave's secure and standard execution modes (\texttt{OCALLs}).
This is particularly evident during operations that initialise sockets and manage peer communication in the TLS handshake.
For instance, using a non-blocking socket API requires constant polling when awaiting client responses.
Although such overhead is also present without SGX, it is accentuated due to the enclave transitions.
We point out that the data remains confidential upon leaving the enclave, as the TLS protocol operates within a secure environment.
More generally, SGX also performs slower than standard execution due to security mechanisms introduced in the microcode~\cite{DBLP:conf/netys/VaucherSF18}.

\Cref{fig:case-1}c explores the impact of incorporating broker attestation evidence in the TLS handshake.
The overhead associated with \wasmSGX (using ECDH) with attestation stands at \benchmarkNewConnectionsWithAttestationECDH, whereas \wasmSGX (using PSK) shows a \benchmarkNewConnectionsWithAttestationPSK overhead when compared to their counterparts without attestation.
Notably, the system saturates when exceeding 15 new ECDH connections per second, while PSK performance remains stable.
The added latency is primarily caused by the asymmetric operations involved in evidence generation and signing. 
To enhance scalability and manage this overhead over a larger set of peers, Mosquitto offers a feature called \emph{broker bridge} that can distribute the workload across multiple brokers, although an in-depth analysis of this feature is left for future work.

\subsection{Messages throughput}
\label{sec:eval-messages}

\begin{figure}[!t]
    \centering
    \includegraphics{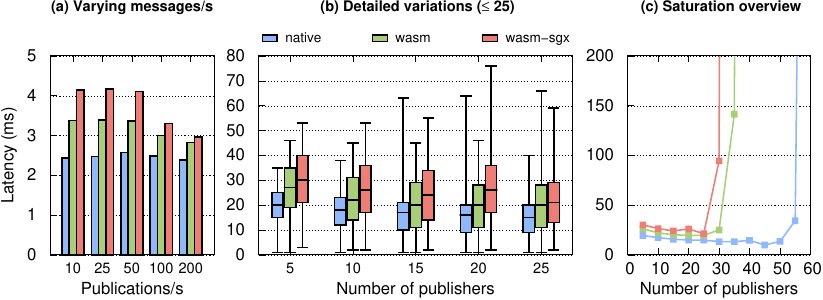}
    \caption{\emph{(a)} shows latency with varying messages per second, \emph{(b)} and \emph{(c)} depict latency as publisher count grows.}
    \label{fig:message-lat}
\end{figure}

In \Cref{fig:message-lat}a, we evaluate the system's throughput by focusing on message delivery scalability through a specific broker.
More specifically, we measure the average latency of delivering a set amount of messages per second (plotted on the x-axis) with a single publisher and subscriber.
The resulting latencies are reported on the y-axis.
Each test case spans 60 seconds and includes a \qty{16}{\kilo\byte} payload of random data.

Latency for the native variant demonstrates a slight rise at a rate of 50 messages per second, followed by a minor decline at 200 messages per second.
In contrast, we note a more pronounced decrease in latency for the \wasm and \wasmSGX variants as the rate of publications increases.
To understand this behaviour, we instrumented the \wasm runtime to monitor the enclave's outgoing calls (\texttt{OCALLs}) and analysed the invoked POSIX functions.
Enclave transitions typically incur performance costs, such as the intrinsic SGX transition and the copy of message buffers into secure memory.
We discovered that POSIX functions related to network message transactions (\ie, \texttt{recvfrom} and \texttt{send}) were called less frequently as the messages-per-second rate increased.
For instance, at a rate of 10 messages per second, the frequencies of \texttt{recvfrom} and \texttt{send} are \num{2601} and \num{1411}, respectively, amounting to an average of 260 and 141 per individual message.
Conversely, at a higher rate of 200 messages per second, these frequencies adjust to \num{49156} and \num{23498}, respectively, meaning a smaller average of 246 and 117 per message.
As these observations appear to be caused by the behaviour of Mosquitto's broker itself, we did not conduct additional investigations.

Overall, the \wasm and \wasmSGX variants show performance slowdown factors of \benchmarkThroughputWasmVsNative and \benchmarkThroughputWasmSGXVsNative, respectively.
Despite these factors, the system scales well and offers enhanced security and portability benefits.

\subsection{Scaling the publishers}
\label{sec:eval-publishers}

As our final experiment, we assess system scalability by increasing the number of publishers plotted on the x-axis, to observe the impact on message delivery latency, shown on the y-axis.
The number of subscribers is held constant at 25, with each publisher sending 5 messages per second.
Each test case lasts 60 seconds and uses a payload of \qty{16}{\kilo\byte} of random data.

\Cref{fig:message-lat}b reveals that all the variants follow a similar trend.
Latency generally remains stable but decreases slightly as the number of publishers increases.
The slowdown of \wasm compared to native execution is \benchmarkScalingPublishersWasmVsNative, while for \wasmSGX against native variant is \benchmarkScalingPublishersWasmSGXVsNative.
Our analysis further investigates the point at which the system begins to saturate, thereby affecting its responsiveness.
\Cref{fig:message-lat}c highlights this limit for each variant, indicating system saturation when exceeding 25 publishers for \wasmSGX, followed by \wasm at 30 publishers.
The native variant only reaches its limit beyond 55 publishers.
Similarly to the first experiment, we could use Mosquitto's broker bridge feature to handle more publishers. This would increase the overall capacity and reduce the load of individual brokers.

\section{Conclusion}\label{sec:conclusion}
Recent evolution in TEEs by leading CPU manufacturers highlights the growing trend in executing software within untrusted environments while processing increasingly sensitive data.
The \pubsub model stands out as an effective mechanism to scale and distribute computations across varied architectures, and \wasm emerges as a suitable common environment for such tasks.
However, a gap remains in establishing standardised protocols and tools that leverage the rapid research breakthroughs in trusted execution within the cloud-edge continuum.

Addressing this, we propose a secure and attested \pubsub system compatible with most of the state-of-the-art TEEs.
We achieve this by using \wasm and trusted runtimes running in these TEEs.
To incorporate mutual attestation in network communications, we suggest enhancing the TLS protocol.
This modification embeds attestation evidence through extension points in the TLS handshake, preserving the original standard.

Our evaluation demonstrates that the security and portability improvements introduced by our approach effectively balance the additional overheads, mostly related to the TEEs.
Moreover, leveraging Mosquitto's capability to distribute brokers among peers can further optimise the system, leading to a scalable and secure architecture suitable for large and real-world applications.
Our implementation is freely distributed as an open-source project~\cite{Menetrey2023Opodis}.

\end{document}